\documentclass[journal=jacsat,manuscript=communication]{achemso}

\usepackage{chemformula} % Formula subscripts using \ch{}
\usepackage[T1]{fontenc} % Use modern font encodings
\usepackage{natbib}
\usepackage{graphicx,grffile}
\usepackage[caption=false,lofdepth,lotdepth]{subfig}
\usepackage{dcolumn}
\usepackage{bm}% bold math
\usepackage{amsmath}

\author{Bj\o rn Hafskjold}
\email{bjorn.hafskjold@ntnu.no}
\author{Dick Bedeaux}
\author{Signe Kjelstrup}
\affiliation
{PoreLab, Department of Chemistry,\\ Norwegian University of Science and Technology - NTNU,\\
 Trondheim, Norway}

\author{\O ivind Wilhelmsen}
\affiliation[]{Porelab, SINTEF Energy Research, Trondheim, Norway}

\title{Nonequilibrium thermodynamics of surfaces captures the energy conversions in a shock wave}

\begin{document}

\begin{abstract}

The local entropy production in a shock wave was analysed in the framework of non-equilibrium thermodynamics (NET) of surfaces.
We show that the thermodynamic state variables in the shock front are equal to their equilibrium values, despite lack of global equilibrium in the dense gas.
This observation was used to develop a theory for the entropy production in a shock wave using Gibbs' surface excess properties.
The theoretical results were compared with a numerical evaluation of the entropy balance for the shock front and confirmed by non-equilibrium molecular dynamics (NEMD) simulations.
The NET analysis shows that the dominant contribution to the entropy production is the dissipation of kinetic and compression energy. This opens the door to accurate representations of energy conversions in shock waves. 
\end{abstract}

The basic theory for shock waves was developed in the late 19th century by Rankine and Hugoniot \cite{rankine1870, hugoniot1887}.
Important theoretical developments were made  during and after the second world war \cite{Friedlander1946}, and by use of computer simulations in more recent years \cite {holian1995}, so shock waves are now pretty well understood.
There are, however, remaining questions, such as exactly how the kinetic and compression energy carried by a shock wave is dissipated or converted to other forms.
This is an important question in fields like detonations \cite{kinney2013}, material science \cite{Zhao2017}, and formation and collapse of bubbles \cite{Pecha2000}, to mention a few.
Energy conversion is a topic of thermodynamics, and since shock waves are irreversible processes, more specifically non-equilibrium thermodynamics (NET).
A shock front has a sharp gradient in the density, similar to a liquid-vapor surface.
This has led us to consider the shock front as a surface and use NET for surfaces \cite{kjelstrup2008} as a tool to extract detailed information about the shock wave.
We show that both energy dissipation and reversible conversion can be determined from this analysis and that it gives new information about energy conversion at the shock front.
We start by deriving the governing equations for the Gibbs excess method applied to a shock wave.

\textit{The Gibbs excess method.} We consider a planar shock wave moving in positive $x$-direction (from left to right).
The shock front is treated as a discontinuity represented by excess variables for the surface (in excess of the bulk phase).
This is similar to the typical treatment of \textit{e.g.} vapor-liquid interfaces.
For example, the surface excess mass density is defined by
\begin{equation}
\rho^{\text{s}}=\int_{x^\text{d}}^{x^\text{u}} \rho(x) 
-\left [\rho^{\text{d}}(x)\Theta
(\ell -x)+\rho^{\text{u}}(x)\Theta (x-\ell ) \right ] dx,
\label{eqn:D}
\end{equation}
where superscript "s" denotes a surface excess property, $\ell $ is the position of the surface, $\Theta$ is the Heaviside step function, and $x^\text{d} < \ell$ and $x^\text{u}>\ell$ are positions in the bulk phases.
The superscripts "d" and "u" denote the extrapolated values of $\rho(x)$ from the bulk values on the downstream (left) and upstream (right) side of the shock.
Eq. \eqref{eqn:D} is the Gibbs definition of excess densities \cite{Gibbs1961}.
Furthermore, we assume that thermodynamic relations between surface variables remain valid also when the system at large is out of equilibrium, as introduced by Bedeaux, Albano and Mazur \cite{Bedeaux1976, Albano1987}.
Many theoretical and simulation studies have showed that this assumption applies to interfaces perturbed far beyond global equilibrium \cite{Savin_2012,kjelstrup2008,de2006}.

In the Gibbs excess method, one must define a dividing surface.
We will do this by requiring that $\rho^{\text{s}}$ equals zero, which determines $\ell(t)$.
The surface moves with a velocity given by
\begin{equation}
v^{\text{s}}(t)=\frac{d\ell (t)}{dt}.
\label{eqn:B.3}
\end{equation}
The entropy density is represented as \cite{kjelstrup2008}
\begin{equation}
\rho_\text{s}(x,t)=\rho_\text{s}^{\text{d}}(x,t)\Theta (\ell -x) +\rho_\text{s}^{\text{s}}(t)\delta (x-\ell ) +\rho_\text{s}^{\text{u}}(x,t)\Theta (x-\ell )  
\label{eqn:F}
\end{equation}
where $\rho_\text{s}^\text{s}$ is the surface excess entropy density.

The balance equation for entropy for the planar shock wave is
\begin{equation}
\frac{\partial}{\partial t} \rho_\text{s}(x,t)+\frac{\partial }{\partial x}%
J_\text{s}(x,t)=\sigma(x,t),
\label{eqn:A}
\end{equation}
where $\rho_\text{s}$, $J_\text{s}$, and $\sigma$ are the entropy density, entropy flux, and entropy production per unit volume, respectively.
The entropy flux in Eq. \eqref{eqn:A} is:
\begin{equation}
J_\text{s}(x,t)=\frac{J'_\text{q}(x,t)}{T(x,t)} + \rho_\text{s}(x,t)v_{\text{cm},x}(x,t)
\label{eqn:B}
\end{equation}
where $J_\text{q}'$ is the measurable heat flux, $T$ is the temperature, and $v_{\text{cm},x}$ is the local streaming velocity.

\begin{figure*}[!ht]
\centering
  \subfloat[]{\includegraphics[width=0.8\columnwidth]{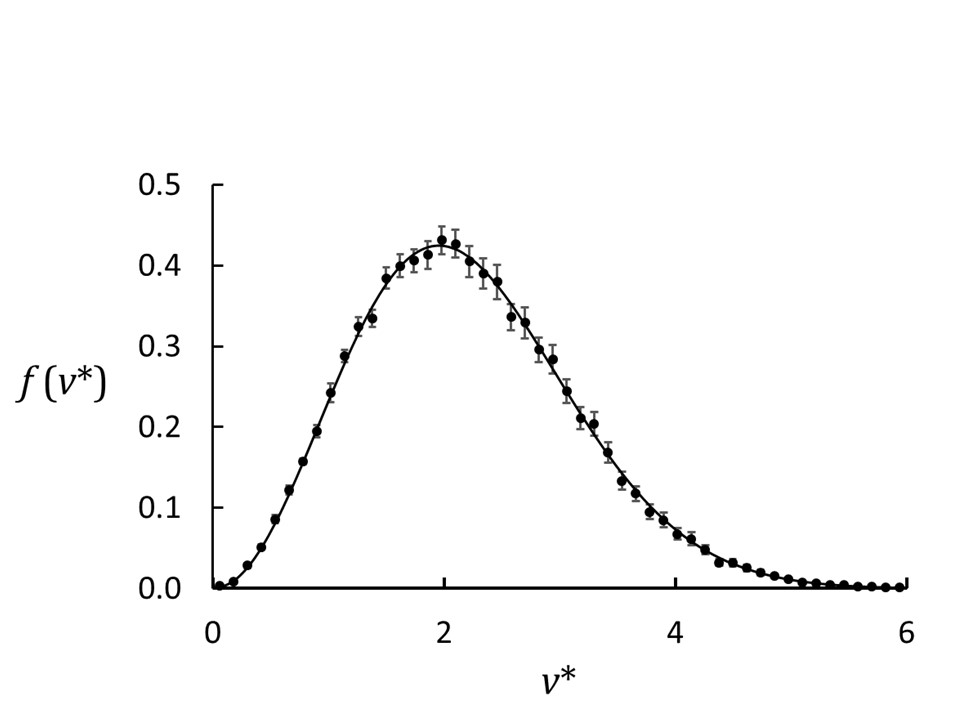}
    \label{fig:maxwell}}
    \qquad
  \subfloat[]{
  \includegraphics[width=0.8\columnwidth]{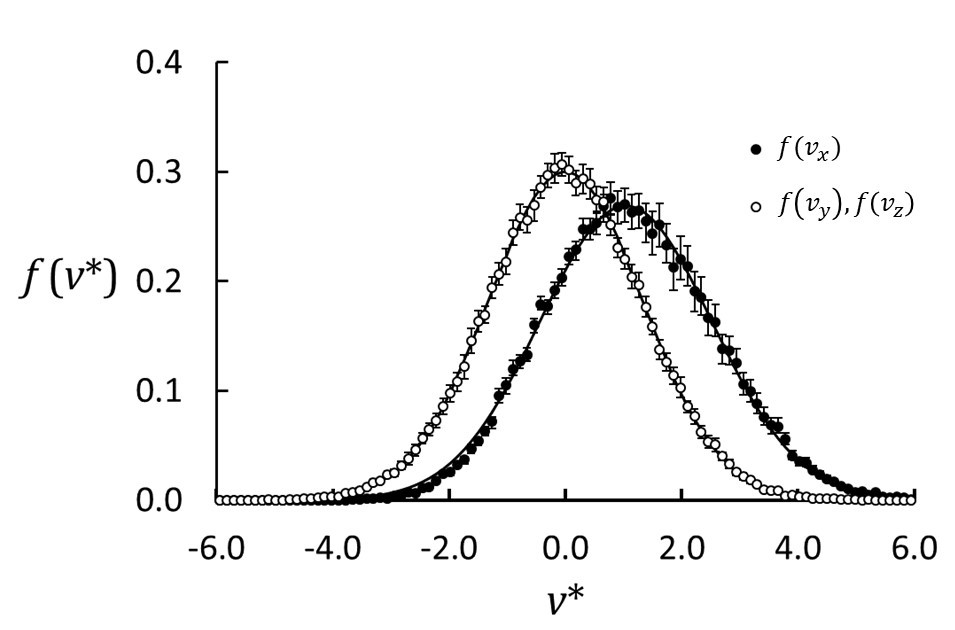}
  \label{fig:gaussian}}
\caption{(a) Particle speed distribution  from NEMD (dots) and a fitted Maxwell-Boltzmann distribution (line) at time $t^*=1000$.
The error bars represent one standard error.
The streaming velocity in $x$-direction was subtracted from the $v_x$-component prior to computing the speed. (b)  Particle velocity distributions in longitudinal and transverse directions for the same condition as in (a).
The difference in mean values corresponds to the streaming velocity in $x$-direction.
The difference in distribution variances corresponds to the difference in longitudinal and transverse temperatures.}
\label{fig:tempz}
\end{figure*}
Substituting Eq. \eqref{eqn:F} into Eq. \eqref{eqn:A}, we obtain after some algebra the following balance equation for the surface excess entropy density, \textit{i.e.} the entropy that is assigned to the shock front as represented by excess and extrapolated variables:
\begin{equation}
\frac{d\rho_\text{s}^{\text{s}}}{dt}+\left[ J_\text{s}-v^{\text{s}}\rho_\text{s}\right]
_{-}=\sigma^{\text{s}}(t) 
\label{eqn:G}
\end{equation}
where we have used the notation 
\begin{equation}
\left[ J_\text{s}-v^{\text{s}}\rho_\text{s}\right] _{-}\equiv J_\text{s}^{\text{u}}-v^{\text{s}}\rho_\text{s}^{\text{u}}-J_\text{s}^\text{d}+v^\text{s}\rho_\text{s}^{\text{d}}  
\label{eqn:H}
\end{equation}
for the difference across the surface.
Here, $J_\text{s}-v^{\text{s}}\rho_\text{s}=J_q' /T+\rho_\text{s}(v-v^{\text{s}})$ is the entropy flux in the surface frame of reference.
The excess entropy density, $\rho_\text{s}^\text{s}$, is found by replacing the mass density with the entropy density in Eq. \eqref{eqn:D} and using the same value of $\ell$ as determined by the Gibbs construction for $\rho^\text{s}$.

We introduce next the Gibbs equation applied to the excess densities of the surface,
\begin{equation}
d \rho_\text{u}^{\text{s}}=T^{\text{s}}d \rho_\text{s}^{\text{s}}+ \mu^{\text{s}} d \rho^{\text{s}} 
\label{eqn:I}
\end{equation}
where $\rho_\text{u}^\text{s}$ is the excess internal energy density, the surface temperature is defined as $T^\text{s}=\partial \rho_\text{u}^{\text{s}}/\partial\rho_\text{s}^{\text{s}}$ at constant $\rho^\text{s}$, and 
$\mu^{\text{s}}$ is the specific Gibbs energy of the surface.
Note that $\rho^\text{s}$ in Eq. \eqref{eqn:I} equals zero by construction.
The $\rho_\text{u}^{\text{s}}$ was found in the same way as $\rho_\text{s}^{\text{s}}$, \textit{i.e.} by replacing $\rho$ by $\rho_\text{u}$ in Eq.~\eqref{eqn:D}.
The local equilibrium hypothesis in the excess description amounts to assuming that Eq.~\eqref{eqn:I} is valid \cite{de2006}.
Rearranging Eq.~\eqref{eqn:I} gives
\begin{equation}
\frac{d\rho_\text{s}^{\text{s}}}{dt}=\frac{1}{T^{\text{s}}}\frac{d \rho_\text{u}^{\text{s}}}{dt}.
\label{eqn:L}
\end{equation}
Conservation of energy across the shock leads to the following balance equation for the excess internal energy density:
\begin{equation}
\frac{d \rho_\text{u}^{\text{s}}}{dt}+\left[ \left\{ \rho_\text{u} +  \textsf{P}_{xx}+\frac{1}{2}\rho \left( v-v^{\text{s}}\right )
^{2}\right\} \left( v-v^{\text{s}}\right ) +J_{q}^\prime \right] _{-} =0
\label{eqn:C.7}
\end{equation}
where $\textsf{P}_{xx}=p+\Pi_{xx}$ is the $xx$-component of the pressure tensor (including the viscous pressure component $\Pi_{xx}=-(\frac{4}{3}\eta_\text{S}+\eta_\text{B})\frac{\partial v}{\partial x}$).
All properties in the brackets are bulk properties. 
By introducing Eq. \eqref{eqn:C.7} into Eq. \eqref{eqn:L} and
comparing the result with the entropy balance, Eq. \eqref{eqn:G}, we obtain the
following expression for the excess entropy production, using the same bracket notation as in Eq. \eqref{eqn:H}:
\begin{equation}
\sigma ^{\text{s}} =[\sigma_q]_- + [ \sigma_j]_-,
\label{eqn:N}
\end{equation}
where 
\begin{equation}
\sigma_q =J_q'\left( \frac{1}{T} - \frac{1}{T^\text{s}}\right)
\label{eqn:Na}
\end{equation}
and
\begin{equation}
\sigma_j =- \frac{j}{T^{
\text{s}}} \left (  \mu + (T-T^\text{s})\frac{\rho_\text{s}}{\rho}+ \frac{\Pi_{xx}}{\rho} +\frac{1}{2} ( v-v^{\text{s}}) ^{2} \right).
\label{eqn:Nb}
\end{equation}
In Eq. \eqref{eqn:Nb}, $\mu$ is the specific Gibbs energy in the bulk and  $j=\rho(v-v^\text{s})$ is the mass flux in the surface frame of reference.

Eqs. \eqref{eqn:N} - \eqref{eqn:Nb} are the key results presented in this work.
They contain thermodynamic properties that are available from the equation of state plus the thermal conductivity and viscosity of the bulk phases.
We will now show how NEMD simulations were used to test Eqs. \eqref{eqn:N} - \eqref{eqn:Nb}.

\begin{figure*}[ht]
\includegraphics[width=0.5\linewidth]{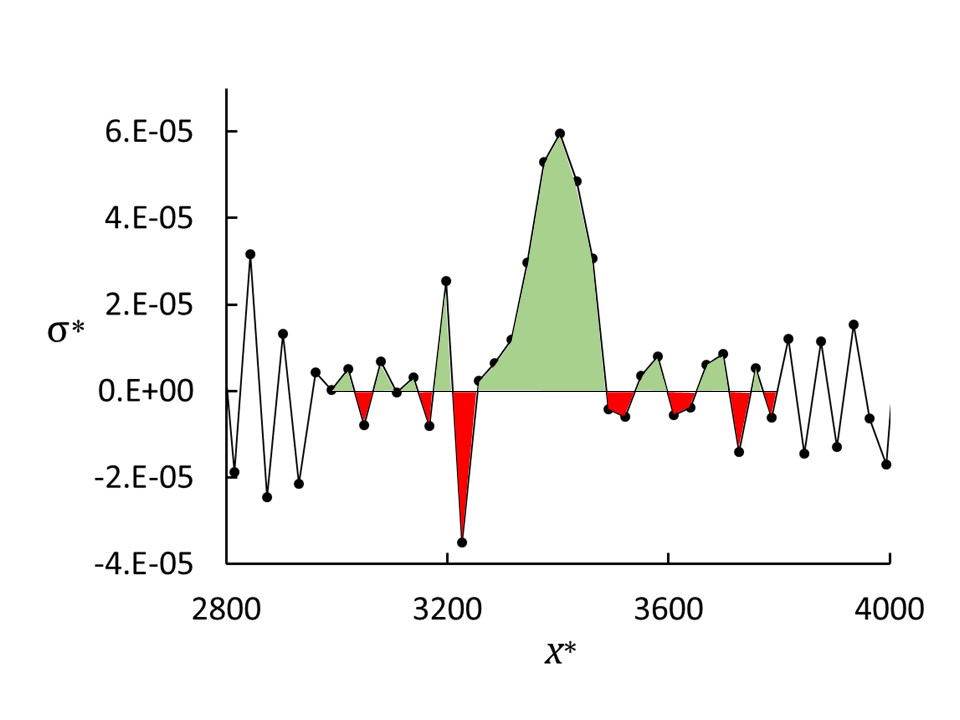}
\label{fig:sigma1}
\caption{The local entropy production, $\sigma(x,y)$ at $t^*=1000$ from Eq. \eqref{eqn:A}.
The colored areas represent the positive (green) and negative (red) contributions to the integral $\sigma^{\text{s}}\approx\int \sigma (x) dx$ from $x^*=3000$ to $x^*=3800$.}
\label{fig:entropyz}
\end{figure*}

\textit{NEMD simulations.}
A planar shock wave was generated by a sudden local heating of an equilibrated one-component system of $N=524,288$ Lennard-Jones/spline particles in a volume $V$ at a reduced density $n^* = N\sigma_{\text{LJ}}^3/V = 0.01$ and reduced temperature $T^* = k_\text{B} T/\epsilon = 1.0$.
The parameters $\sigma_{\text{LJ}}$ and $\epsilon$ are the usual Lennard-Jones parameters and $k_\text{B}$ is Boltzmann's constant.
The blast was initiated at time $t^*= \frac{t}{\sigma_{\text{LJ}}}\left ( \frac{\epsilon}{m}\right )^{1/2}=0$, where $m$ is the particle mass, and at position $x^* = x/\sigma_{\text{LJ}} = 0$.
The subsequent density-, temperature-, and pressure profiles were computed in the \textit{NVE} ensemble as functions of time by dividing the system into control volumes (layers) of thickness $\Delta x^* = 29.5$ and time slots of length $t^*=10$.
The mean free path of the particles upstream of the shock was $\lambda^* = 1/(\sqrt{2} \pi n^*)=22.5$ as determined by elementary kinetic theory.
Averages and uncertainties were based on 20 independent runs starting from randomized equilibrium configurations. 

The speed of sound in the gas upstream of the shock was determined from independent equilibrium molecular dynamics simulations and found to be 1.298, which is very close to the ideal-gas value of 1.291. The blast caused the shock wave to travel at a slowly retarding supersonic speed with a Mach number of 2.1. 
This is a weak shock, on the borderline of the validity range of the local equilibrium condition~\cite{margolin2017, margolin2017b}.

\textit{The question of local equilibrium.} Shock waves are non-equilibrium structures.
For instance, the velocity distribution and the kinetic temperature in the shock front is anisotropic \cite{Holian2010,Hoover2014}.
However, many studies have confirmed that the classical local equilibrium hypothesis \cite{kjelstrup2008} holds when the interfacial properties are described by Gibbs excess variables~\cite{kjelstrup2008,de2006,Savin_2012}.
In agreement with previous results \cite{holian1993,Hoover2014}, we also found that the local kinetic temperature was anisotropic in the shock front.
On the other hand, we found that the Boltzmann $H$-function based on the particle speeds from the NEMD data was consistent with a state of \textit{local} equilibrium.
This is illustrated in Fig.~\ref{fig:maxwell}, based on the speed of 35,996 particles (total from 20 runs) that were in the control volume of thickness $\Delta x^*$, centred at the shock wave front at $x^* = 3420$, at $t^* = 1000$.
The fitted Maxwell-Boltzmann distribution gave a temperature $T^* = 1.9 \pm 0.1$, in good agreement with the kinetic temperature $T^*=1.87 \pm 0.03$ (uncertainties given as three standard errors of the mean).
The separate particle velocities in $x$-, $y$- and $z$- directions confirmed the equilibrium longitudinal and transverse distributions and the corresponding local kinetic temperatures.
This is illustrated in Fig. \ref{fig:gaussian}.
We conclude from this that the non-equilibrium entropy determined from the $H$-function agrees with the equilibrium value within the estimated uncertainty.

\textit{Direct numerical evaluation of the entropy balance equation.}
In order to verify the validity of Eqs. \eqref{eqn:N} - \eqref{eqn:Nb} and the Gibbs excess method, we computed the local entropy production by direct numerical evaluation of the entropy balance equation, Eq.~\eqref{eqn:A}, over the surface region.
The only assumption behind this method is that the local properties are determined by the equation of state as discussed above.
The first term in Eq. \eqref{eqn:A}, $(\partial \rho_\text{s} / \partial t)$ was evaluated by numerical differentiation of the data from NEMD using a five-point method.

The heat flux $J_q'$ in Eq. \eqref{eqn:B} was computed directly from the NEMD simulations.
Although the heat flux has a sharp peak in the shock front, it contributes at most only 3~\% to $J_\text{s}$ in the present case.
The second term in Eq. \eqref{eqn:A}, $(\partial J_s/\partial x)$, was computed by numerical differentiation of the entropy flux profiles.
The two terms on the left-hand side of Eq. \eqref{eqn:A} are large and of opposite sign around the shock front.
The uncertainty in the sum at the left-hand-side of Eq.~\eqref{eqn:A} is therefore large.
Nevertheless, the entropy production shown in Fig.~\ref{fig:entropyz} displays a distinct positive peak around the shock front, in agreement with the second law of thermodynamics.

Finally, $\sigma(x)$ was integrated from $x^*=3000$ to $x^*=3800$ with a simple trapezoidal rule.
Noise on both sides of the peak gives positive (shaded green in Fig.~\ref{fig:entropyz}) and negative (shaded red) contributions to the integral, which cancel out to zero.
The non-zero contribution to the spacial integral of $\sigma(x,t)$ is from the peak centred on the shock front.
The integral determines the surface excess entropy production per cross sectional area, $\sigma^{\text{s}}$.
An estimate at $t^* =1000$ gave an excess entropy production in the shock wave $\sigma^{\text{s}*} = 0.007\pm 0.002$.

\begin{figure*}[ht]
  \subfloat[][]{ \centering
    \includegraphics[width=0.9\columnwidth]{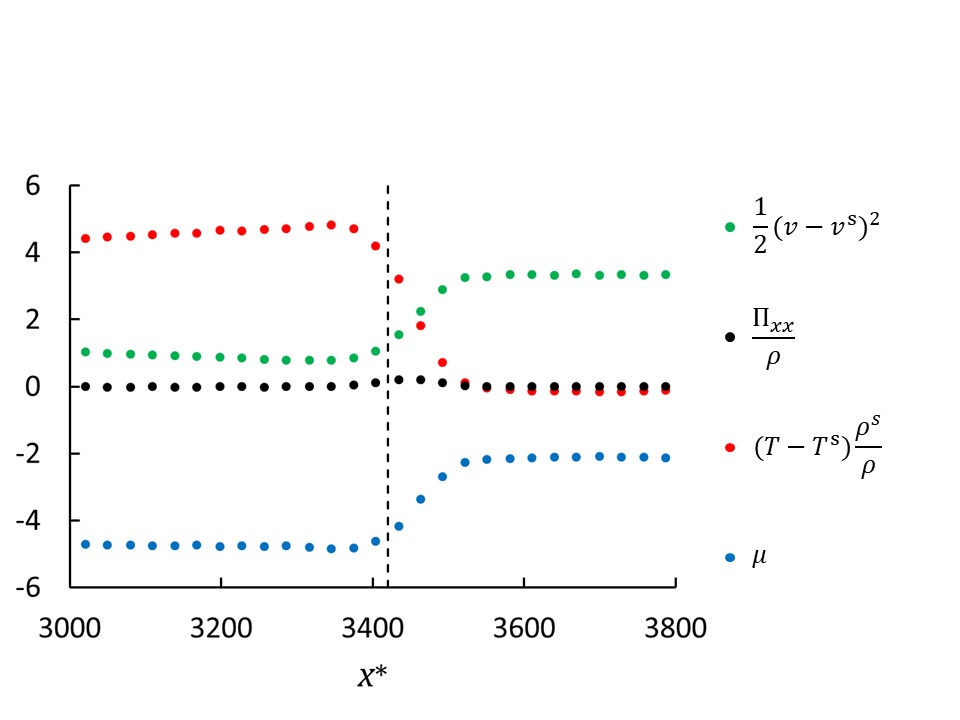}
    \label{fig:threeterms}}
  \subfloat[][]{ \centering
  \includegraphics[width=0.9\columnwidth]{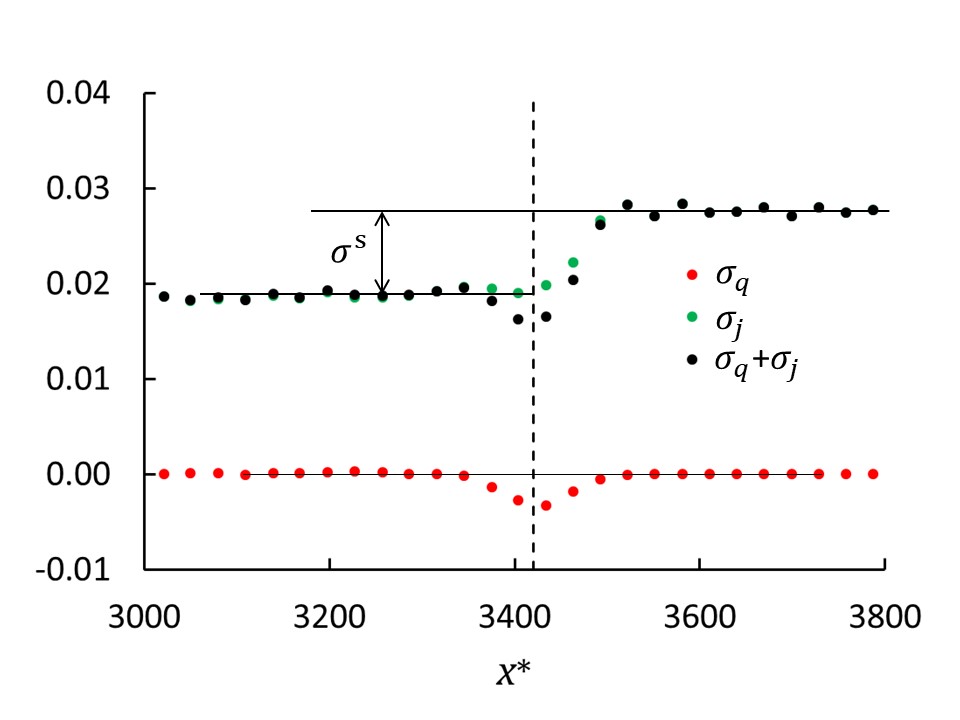}
  \label{fig:sigmasurface}}
\caption{(a) Comparison of the four contributions to $\sigma_j$ defined in Eq.~\eqref{eqn:Nb} at $t^*=1000$. The sum of these functions are extrapolated to the surface as described below Eq. \eqref{eqn:D}. (b) The two terms in Eq. \eqref{eqn:N} and their sum at $t^*=1000$. The surface excess entropy production is the difference between the extrapolated values from right and from left as indicated by the double arrow. The vertical dotted line marks the surface position.}
\label{fig:surfacez}
\end{figure*}

\textit{Numerical evaluation of Eqs. \eqref{eqn:N} - \eqref{eqn:Nb}.}
%NEMD simulations were used to provide 
%Eqs. \eqref{eqn:N} - \eqref{eqn:Nb} will now be used to analyse the various contributions to the entropy production.
The dominant contribution to the excess entropy production comes from $\sigma_j$.
The term $\sigma_q$ is small because the heat flux in the bulk phases is small (slightly negative downstream of the front and zero upstream).
Fig.~\ref{fig:threeterms} shows the profiles of the four terms in the bracket in Eq. \eqref{eqn:Nb} at $t^*=1000$.
The viscous  pressure term varies little over the shock front.
The difference between the extrapolated values is practically zero.
The kinetic energy term includes the center-of-mass velocity relative to the shock wave velocity. This relative velocity is larger upstream than downstream, so the difference defined by the bracket is positive.
The specific Gibbs energy  is the difference between the specific enthalpy and the product of temperature and specific entropy, $\mu=h-Ts$.
Both $h$ and $s$ increase when the shock wave passes, but the entropy term increases more than the enthalpy, so the total effect is a decrease in the specific Gibbs energy.
The term $(T-T^\text{s})\rho_\text{s}/\rho$ increases because both $(T-T^\text{s})$ and $\rho_\text{s}/\rho$ increase from upstream to downstream.
The mass flux is constant across the shock front because mass is conserved, and therefore equal to the upstream value, $j=-\rho v^\text{s}$.
In total, the term $\sigma_j$ is positive.
Hence, for the propagating shock examined in this work, the overall picture is that kinetic energy and chemical energy are partially converted to entropy across the shock front.

The two contributions to the entropy production and their sum are shown in Fig.~\ref{fig:sigmasurface}.
The relatively small contribution from heat conduction and viscous dissipation is a consequence of the low density of the gas, and we expect these terms to be larger in fluids with higher densities.
Eqs.~\eqref{eqn:N}~-~\eqref{eqn:Nb} provide a tool to quantify this.

The total excess entropy production as given by Eq.~\eqref{eqn:N} is a difference between properties extrapolated to the surface position.
This extrapolation is illustrated by the horizontal lines in Fig.~\ref{fig:sigmasurface} and the difference is illustrated by the double arrow.
We emphasize that the values in the shock-front region have no significance in this context, only the extrapolated values are relevant.
We found that the total excess entropy production per unit surface area was $\sigma^{\text{s}*}=0.009 \pm 0.001$, which compares well with the value for the excess entropy production based on Eq. \eqref{eqn:A}.

\textit{Conclusions.} We have presented a new method to analyze the entropy production in a propagating shock wave by use of non-equilibrium thermodynamics for surfaces, using surface excess variables.
The only assumptions behind this method is that the local properties are determined by the equation of state and that the Gibbs equation holds for the surface excess properties.
These assumptions have been found valid for surfaces, and so also in the present case.
A numerical evaluation was made with data from non-equilibrium molecular dynamics simulations of a weak shock wave.
Within the accuracy of the simulations, this method gave the same surface excess entropy production as a direct numerical evaluation of the local entropy balance in the shock-front region.
The new method is a powerful tool for analysis of energy conversions in shock waves because it quantifies the different contributions to the excess entropy production.
A consistent representation of dissipation in shocks is of key importance for the dynamic description of shock waves in a variety of fields.

\textit{Acknowledgements.}
The NEMD simulations were performed on resources provided by UNINETT Sigma2 - the National Infrastructure for High Performance Computing and
Data Storage in Norway and by Department of Chemistry at The Norwegian University of Science and Technology - NTNU.

\bibliography{shockwave}

\providecommand{\latin}[1]{#1}
\makeatletter
\providecommand{\doi}
  {\begingroup\let\do\@makeother\dospecials
  \catcode`\{=1 \catcode`\}=2 \doi@aux}
\providecommand{\doi@aux}[1]{\endgroup\texttt{#1}}
\makeatother
\providecommand*\mcitethebibliography{\thebibliography}
\csname @ifundefined\endcsname{endmcitethebibliography}
  {\let\endmcitethebibliography\endthebibliography}{}
\begin{mcitethebibliography}{19}
\providecommand*\natexlab[1]{#1}
\providecommand*\mciteSetBstSublistMode[1]{}
\providecommand*\mciteSetBstMaxWidthForm[2]{}
\providecommand*\mciteBstWouldAddEndPuncttrue
  {\def\EndOfBibitem{\unskip.}}
\providecommand*\mciteBstWouldAddEndPunctfalse
  {\let\EndOfBibitem\relax}
\providecommand*\mciteSetBstMidEndSepPunct[3]{}
\providecommand*\mciteSetBstSublistLabelBeginEnd[3]{}
\providecommand*\EndOfBibitem{}
\mciteSetBstSublistMode{f}
\mciteSetBstMaxWidthForm{subitem}{(\alph{mcitesubitemcount})}
\mciteSetBstSublistLabelBeginEnd
  {\mcitemaxwidthsubitemform\space}
  {\relax}
  {\relax}

\bibitem[Rankine \latin{et~al.}(1870)Rankine, \latin{et~al.}
  others]{rankine1870}
Rankine,~W. J.~M., \latin{et~al.}  XV. On the thermodynamic theory of waves of
  finite longitudinal disturbance. \emph{Philosophical Transactions of the
  Royal Society of London} \textbf{1870}, \emph{160}, 277--288\relax
\mciteBstWouldAddEndPuncttrue
\mciteSetBstMidEndSepPunct{\mcitedefaultmidpunct}
{\mcitedefaultendpunct}{\mcitedefaultseppunct}\relax
\EndOfBibitem
\bibitem[Hugoniot(1887)]{hugoniot1887}
Hugoniot,~H. Memoir on the propagation of movements in bodies, especially
  perfect gases (first part). \emph{J. de l’Ecole Polytechnique}
  \textbf{1887}, \emph{57}, 3--97\relax
\mciteBstWouldAddEndPuncttrue
\mciteSetBstMidEndSepPunct{\mcitedefaultmidpunct}
{\mcitedefaultendpunct}{\mcitedefaultseppunct}\relax
\EndOfBibitem
\bibitem[Friedlander(1946)]{Friedlander1946}
Friedlander,~F.~G. \emph{Proc. Roy. Soc. London} \textbf{1946}, \emph{A186},
  322 -- 344\relax
\mciteBstWouldAddEndPuncttrue
\mciteSetBstMidEndSepPunct{\mcitedefaultmidpunct}
{\mcitedefaultendpunct}{\mcitedefaultseppunct}\relax
\EndOfBibitem
\bibitem[Holian(1995)]{holian1995}
Holian,~B. Atomistic computer simulations of shock waves. \emph{Shock waves}
  \textbf{1995}, \emph{5}, 149--157\relax
\mciteBstWouldAddEndPuncttrue
\mciteSetBstMidEndSepPunct{\mcitedefaultmidpunct}
{\mcitedefaultendpunct}{\mcitedefaultseppunct}\relax
\EndOfBibitem
\bibitem[Kinney and Graham(2013)Kinney, and Graham]{kinney2013}
Kinney,~G.~F.; Graham,~K.~J. \emph{Explosive shocks in air}; Springer Science
  \& Business Media, 2013\relax
\mciteBstWouldAddEndPuncttrue
\mciteSetBstMidEndSepPunct{\mcitedefaultmidpunct}
{\mcitedefaultendpunct}{\mcitedefaultseppunct}\relax
\EndOfBibitem
\bibitem[Zhao \latin{et~al.}(2017)Zhao, Kad, Wehrenberg, Remington, Hahn, More,
  and Meyers]{Zhao2017}
Zhao,~S.; Kad,~B.; Wehrenberg,~C.~E.; Remington,~B.~A.; Hahn,~E.~N.;
  More,~K.~L.; Meyers,~M.~A. Generating gradient germanium nanostructures by
  shock-induced amorphization and crystallization. \emph{Proceedings of the
  National Academy of Sciences} \textbf{2017}, \emph{114}, 9791--9796\relax
\mciteBstWouldAddEndPuncttrue
\mciteSetBstMidEndSepPunct{\mcitedefaultmidpunct}
{\mcitedefaultendpunct}{\mcitedefaultseppunct}\relax
\EndOfBibitem
\bibitem[Pecha and Gompf(2000)Pecha, and Gompf]{Pecha2000}
Pecha,~R.; Gompf,~B. Microimplosions: Cavitation Collapse and Shock Wave
  Emission on a Nanosecond Time Scale. \emph{Phys. Rev. Lett.} \textbf{2000},
  \emph{84}, 1328--1330\relax
\mciteBstWouldAddEndPuncttrue
\mciteSetBstMidEndSepPunct{\mcitedefaultmidpunct}
{\mcitedefaultendpunct}{\mcitedefaultseppunct}\relax
\EndOfBibitem
\bibitem[Kjelstrup and Bedeaux(2008)Kjelstrup, and Bedeaux]{kjelstrup2008}
Kjelstrup,~S.; Bedeaux,~D. \emph{Non-Equilibrium Thermodynamics of
  Heterogeneous Systems}; Wiley: Singapore, 2008\relax
\mciteBstWouldAddEndPuncttrue
\mciteSetBstMidEndSepPunct{\mcitedefaultmidpunct}
{\mcitedefaultendpunct}{\mcitedefaultseppunct}\relax
\EndOfBibitem
\bibitem[Gibbs(1961)]{Gibbs1961}
Gibbs,~J.~W. \emph{The Scientific Papers, Vol I: Thermodynamics}; Dover
  Publications, 1961\relax
\mciteBstWouldAddEndPuncttrue
\mciteSetBstMidEndSepPunct{\mcitedefaultmidpunct}
{\mcitedefaultendpunct}{\mcitedefaultseppunct}\relax
\EndOfBibitem
\bibitem[Bedeaux \latin{et~al.}(1976)Bedeaux, Albano, and Mazur]{Bedeaux1976}
Bedeaux,~D.; Albano,~A.~M.; Mazur,~P. Boundary conditions and non-equilibrium
  thermodynamics. \emph{Physica A} \textbf{1976}, \emph{82}, 438--462\relax
\mciteBstWouldAddEndPuncttrue
\mciteSetBstMidEndSepPunct{\mcitedefaultmidpunct}
{\mcitedefaultendpunct}{\mcitedefaultseppunct}\relax
\EndOfBibitem
\bibitem[Albano and Bedeaux(1987)Albano, and Bedeaux]{Albano1987}
Albano,~A.~M.; Bedeaux,~D. Non-equilibrium electro-thermodynamics of
  polarizable multicomponent fluids with an interface. \emph{Physica A}
  \textbf{1987}, \emph{147}, 407--435\relax
\mciteBstWouldAddEndPuncttrue
\mciteSetBstMidEndSepPunct{\mcitedefaultmidpunct}
{\mcitedefaultendpunct}{\mcitedefaultseppunct}\relax
\EndOfBibitem
\bibitem[Savin \latin{et~al.}(2012)Savin, Glavatskiy, Kjelstrup, \"{O}ttinger,
  and Bedeaux]{Savin_2012}
Savin,~T.; Glavatskiy,~K.~S.; Kjelstrup,~S.; \"{O}ttinger,~H.~C.; Bedeaux,~D.
  {Local equilibrium of the Gibbs interface in two-phase systems}. \emph{EPL}
  \textbf{2012}, \emph{97}, 40002\relax
\mciteBstWouldAddEndPuncttrue
\mciteSetBstMidEndSepPunct{\mcitedefaultmidpunct}
{\mcitedefaultendpunct}{\mcitedefaultseppunct}\relax
\EndOfBibitem
\bibitem[De~Zarate and Sengers(2006)De~Zarate, and Sengers]{de2006}
De~Zarate,~J. M.~O.; Sengers,~J.~V. \emph{Hydrodynamic fluctuations in fluids
  and fluid mixtures}; Elsevier, 2006\relax
\mciteBstWouldAddEndPuncttrue
\mciteSetBstMidEndSepPunct{\mcitedefaultmidpunct}
{\mcitedefaultendpunct}{\mcitedefaultseppunct}\relax
\EndOfBibitem
\bibitem[Margolin(2017)]{margolin2017}
Margolin,~L.~G. Nonequilibrium entropy in a shock. \emph{Entropy}
  \textbf{2017}, \emph{19}, 368\relax
\mciteBstWouldAddEndPuncttrue
\mciteSetBstMidEndSepPunct{\mcitedefaultmidpunct}
{\mcitedefaultendpunct}{\mcitedefaultseppunct}\relax
\EndOfBibitem
\bibitem[Margolin \latin{et~al.}(2017)Margolin, Reisner, and
  Jordan]{margolin2017b}
Margolin,~L.~G.; Reisner,~J.~M.; Jordan,~P.~M. Entropy in self-similar shock
  profiles. \emph{International Journal of Non-Linear Mechanics} \textbf{2017},
  \emph{95}, 333--346\relax
\mciteBstWouldAddEndPuncttrue
\mciteSetBstMidEndSepPunct{\mcitedefaultmidpunct}
{\mcitedefaultendpunct}{\mcitedefaultseppunct}\relax
\EndOfBibitem
\bibitem[Holian and Mareschal(2010)Holian, and Mareschal]{Holian2010}
Holian,~B.~L.; Mareschal,~M. Heat-flow equation motivated by the ideal-gas
  shock wave. \emph{Physical Review E} \textbf{2010}, \emph{82}, 026707\relax
\mciteBstWouldAddEndPuncttrue
\mciteSetBstMidEndSepPunct{\mcitedefaultmidpunct}
{\mcitedefaultendpunct}{\mcitedefaultseppunct}\relax
\EndOfBibitem
\bibitem[Hoover \latin{et~al.}(2014)Hoover, Hoover, and Travis]{Hoover2014}
Hoover,~W.~G.; Hoover,~C.~G.; Travis,~K.~P. Shock-Wave Compression and
  {J}oule-{T}homson Expansion. \emph{Phys. Rev. Lett.} \textbf{2014},
  \emph{112}, 144504\relax
\mciteBstWouldAddEndPuncttrue
\mciteSetBstMidEndSepPunct{\mcitedefaultmidpunct}
{\mcitedefaultendpunct}{\mcitedefaultseppunct}\relax
\EndOfBibitem
\bibitem[Holian \latin{et~al.}(1993)Holian, Patterson, Mareschal, and
  Salomons]{holian1993}
Holian,~B.~L.; Patterson,~C.; Mareschal,~M.; Salomons,~E. Modeling shock waves
  in an ideal gas: Going beyond the Navier-Stokes level. \emph{Physical review
  E} \textbf{1993}, \emph{47}, R24\relax
\mciteBstWouldAddEndPuncttrue
\mciteSetBstMidEndSepPunct{\mcitedefaultmidpunct}
{\mcitedefaultendpunct}{\mcitedefaultseppunct}\relax
\EndOfBibitem
\end{mcitethebibliography}

\end{document}